\documentclass[10pt]{llncs}
\usepackage[utf8]{inputenc}
\usepackage[english]{babel}  
\usepackage{latexsym}
\usepackage{amsmath,amssymb}
\usepackage[dvips]{graphicx,psfrag}

\usepackage{url}
\usepackage[ruled,linesnumbered]{algorithm2e}

\renewcommand{\vec}[1]{{\mbox{\boldmath$#1$}}}

\newcommand{\MCS}{{MCS}}

\def\argmax{\mathop{\rm argmax}}

\usepackage{color}

\renewcommand{\vec}[1]{{\mathbf #1}}
\title{Pilot, Rollout and Monte Carlo Tree Search Methods for Job Shop Scheduling}

\author{Thomas Philip Runarsson\inst{1} \and Marc Schoenauer\inst{2,4} \and Mich\`ele Sebag\inst{3,4}}
\institute{{School of Engineering and Natural Sciences, University of Iceland} \and {TAO, INRIA Saclay Île-de-France, Orsay, France} \and {TAO, LRI, UMR CNRS 8623, Orsay, France} \and {Microsoft Research-INRIA Joint Centre, Orsay, France}} 
\begin{document}
\maketitle

\selectlanguage{english}

\begin{abstract}
Greedy heuristics may be attuned by looking ahead for each possible
choice, in an approach called the rollout or Pilot method. These
methods may be seen as meta-heuristics that can enhance (any)
heuristic solution, by repetitively modifying a master solution:
similarly to what is done in game tree search, better choices are
identified using lookahead, based on solutions obtained by repeatedly
using a greedy heuristic. This paper first illustrates how the Pilot
method improves upon some simple well known dispatch heuristics for
the job-shop scheduling problem. The Pilot method is then shown to be
a special case of the more recent Monte Carlo Tree Search (MCTS)
methods: Unlike the Pilot method, MCTS methods use random completion
of partial solutions to identify promising branches of the tree. The
Pilot method and a simple version of MCTS, 
using the $\varepsilon$-greedy exploration
paradigms, are then compared within the same framework, consisting of
300 scheduling problems of varying sizes with fixed-budget of
rollouts. Results demonstrate that MCTS reaches better or same results as 
the Pilot methods in this context.
\end{abstract}

\section{Introduction}\label{sec:Intro}
In quite a few domains related to combinatorial optimization, such as
constraint solving \cite{Hutter}, planning or scheduling
\cite{Burke2010}, software environments have been designed to achieve
good performances in expectation over a given distribution of problem
instances.  Such environments usually rely on a portfolio of
heuristics, leaving the designer with the issue of finding the best
heuristics, or the best heuristics sequence, for his particular
distribution of problem instances.

The simplest solution naturally is to use the default heuristics,
assumedly the best one on average on all problem instances. Another
approach, referred to as {\em Pilot or rollout} method, iteratively
optimizes the option selected at each choice point
\cite{Duin1999,Lagoudakis03}, while sticking to the default heuristics
for other choice points. Yet another approach, referred to as {\em
  Monte-Carlo Tree Search} (MCTS) \cite{Kocsis06} and at the origin of
the best current computer-Go players \cite{Gelly07}, has been proposed
to explore the search space while addressing the exploration {\em
  versus} exploitation dilemma in a principled way; as shown by
\cite{Asmuth2011}, MCTS provides an approximation to optimal Bayes
decision. The MCTS approach, rooted in the multi-Armed bandit (MAB)
setting \cite{Lai-Robbins}, iteratively grows a search tree 
through tree walks. For each tree walk, in each node (choice point) 
the selection of the child node (heuristics) is handled as a MAB problem; 
the search tree thus asymmetrically grows to explore the most promising 
tree paths (the most promising sequences of heuristics).

Whereas increasingly used today in sequential decision making
algorithms including games \cite{Rimmel09,Rolet09}, to our
best knowledge MCTS methods have rarely been used within the framework
of combinatorial optimization, with the recent exception of
\cite{Matsumoto2010}. This paper investigates the application of MCTS
to job-shop scheduling, an NP-hard combinatorial optimization problem.
As each job-shop scheduling problem instance defines a deterministic
optimization problem, the standard {\em upper confidence bound applied to tree} (UCT) framework used in
\cite{Matsumoto2010} does not apply. More precisely, the sought
solution is the one with best payoff (as opposed to the one with best
payoff on average); the MAB problem nested in the UCT thus is a max
$k$-armed bandit problem \cite{Streeter2006,Rimmel09}.  Along this
line, the randomized aspects in UCT must be addressed specifically to
fit deterministic problems. Specifically, a critical difficulty lies
in the randomized default handling of the choice points which are
outside the current search tree (in contrast, these choice points are
dealt with using the default heuristics in the pilot methods). Another
difficulty, shared with most MCTS applications, is to preserve the
exploration/ exploitation trade-off when the problem size increases. A
domain-aware randomized default handling is proposed in this paper,
supporting a MCTS-based scheduling approach called {\em Monte-Carlo
  Tree Scheduling} (\MCS).  \MCS\ is empirically validated, using well
established greedy heuristics and the pilot methods based on these
heuristics as baselines. The empirical evidence shows that Pilot
methods significantly outperform the best-known default heuristics;
\MCS\ significantly outperforms on the Pilot methods for small problem
sizes. For larger problem sizes, however, \MCS\ is dominated by the
best Pilot methods, which is partly explained from the experimental
setting as the computational cost of the Pilot methods is about 4
times higher than that of the \MCS\ one. That is, heuristic scheduling
is more costly than random scheduling.

The paper is organized as follows. Job-shop scheduling is introduced
in Section \ref{sec:scheduling}, together with some basic greedy
algorithms based on domain-specific heuristics called {\em dispatching
  rules}. The generic pilot method is recalled in section
\ref{sec:pilot}. The general MCTS ideas are introduced in section
\ref{sec:MCTS}; its adaptation to the job-shop scheduling problem is
described and an overview of \MCS\ is given. and together with its
application to combinatorial problems, and to the job-shop scheduling
problem.  Section \ref{sec:experiments} is devoted to the empirical
validation of the proposed approach. After describing the experimental
setting, the section reports on the \MCS\ results on different problem
sizes, with the dispatching rules and the Pilot methods as
baselines. The paper concludes with a discussion of these results and
some perspectives for further research.

\section{Job Shop Scheduling and priority dispatching rules}\label{sec:scheduling}

Scheduling is the sequencing of the order in which a set of
\emph{jobs} $j \in J:=\{1,..,n\}$ are processed through a set of
\emph{machines} $a\in M:=\{ 1,..,m_j\}$. In a \emph{job shop}, the order
in which a job is processed through the machines is
predetermined. In a \emph{flow shop} this order is the same for all
jobs, and in a \emph{open shop} the order is arbitrary. We will
consider here only the \emph{job shop}, where the jobs are
strictly-ordered sequences of operations. A
job can only be performed by one type of machine and each machine
processes one job at a time. Once a job is started it must be
completed. The performance metric for scheduling problems is generally based on flow or dues date. Here we will consider the completion time for the last job or the so called makespan.

Each job has a specified processing time $p(j,a)$ and the order through the machines is given by the permutation vector $\sigma$ ($\sigma(j,i)$ is the $i^{th}$ machine for job $j$). Let $x(j,a)$ be the start time for job $j$ on machine $a$, then 
\begin{equation}\label{eq:Permutation}
   x(j,\sigma(j,i)) \geq x(j,\sigma(j,i-1) + p(j,\sigma(j,i-1))\quad j\in\{1,..,n\},\; i\in\{2,..,m_j\}
\end{equation}
The disjunctive condition that each machine can handle at most one job at a time is the following: 
\begin{equation}\label{eq:OneJobPerMac}
   x(j,a) \geq x(k,a)+p(k,a) \quad\textrm{or}\quad  x(k,a)+p(k,a) \geq x(j,a)
\end{equation}
for all $j,k\in J, j\ne k$ and $a\in M$.  
The makespan can then be formally defined as
\begin{equation}
  z = \max\{x(j, \sigma(j,m_j))+p(j,m_j)\;|\;j\in J\}.
\end{equation}

Smaller problems can be solved using a specialized branch and bound procedure \cite{Brucker2007} and an algorithmic implementation may be found as part of LiSA \cite{Lisa2011}. Jobs up to 14 jobs and 14 machines can still be solved efficiently, but at higher dimensions, the problems rapidly become intractable.
Several heuristics have been proposed to solve job shop problems when their size becomes too large for exact methods. One such set of heuristics are based on \emph{dispatch rules}, i.e. rules to decide which job to schedule next based on the current state of all machines and jobs. A survey of over 100 such rules may be found in \cite{Panwalkar1977a}. Commonly used priority
dispatch rules have been compared on a number of benchmark problems
in \cite{Kawai2005}. When considering the makespan as a performance
metric, the rule that selects a job which has the {\em Most WorK
  Remaining} (MWKR, the job with the longest total remaining
processing time) performed overall best. It was followed by the rule
that selects a job with the {\em Shortest Processing Time} (SPT), and
by the rule that selects a job which the {\em Least Operation Number}
(LOPN). These rules are among the simplest ones, and are by no means
optimal. However, only these 3 rules will be considered in the
remaining of this paper. In particular, experimental results of the
corresponding 3 greedy algorithms can be found in Section
\ref{sec:experiments}.

The simplest way to use any of these rules is to embed them in a greedy algorithm: the jobs are processed in the order 
given by the repeated application of the chosen rule. Algorithm \ref{alg:greedy} gives the pseudo-code of such an 
algorithm. The variable $t_j$ represents which machine is next in line for job $j$ (more precisely machine 
$\sigma(j,t_i)$). When starting with an empty schedule, one would set $t_j \leftarrow 1$ for $j \in J$ and ${\cal S} = 
\emptyset$.
At each step of the algorithm, one job is chosen according to the dispatching rule {\bf R} (line \ref{lin:chooseJob}), and the job is scheduled on the next machine in its own list, i.e., the pair (job, machine) is added to the partial schedule $\cal S$ (line \ref{lin:scheduleJob}) ($\oplus$ denotes the concatenation of two lists). 

\begin{algorithm}
\SetAlFnt{\footnotesize}
\SetCommentSty{emph}
\SetKwData{Left}{left}
\SetKwData{This}{this}
\SetKwData{Up}{up}
\SetKwFunction{Rollout}{Rollout}
\SetKwFunction{Simulate}{Simulate}
\SetKwInOut{Input}{input}
\SetKwInOut{Output}{output}
\AlFnt
\Input{Partial sequence ${\cal S}_0$, $\vec{t}=(t_1,\ldots,t_n)$, and heuristic {\bf R}}
\Output{An objective to maximize, for example negative makespan}
\BlankLine
\While{$\exists j\in J \; ; \; t_j < m_j$}{
	$b = \mbox{\bf R}({\cal S},t_j\; ; t_j < m_j)$ \tcp*{Apply {\bf R} to current partial schedule, get next job} 
		\nllabel{lin:chooseJob}
		${\cal S}\leftarrow {\cal S} \oplus \{(b, \sigma(b,t_b))\}$ \tcp*{Schedule job on its next machine}
		\nllabel{lin:scheduleJob}
		$t_b\leftarrow t_b+1$ \tcp*{Point to next machine for job $b$}\nllabel{lin:iterate}
}
\caption{Greedy (Pilot) heuristic }\label{alg:greedy}
\end{algorithm}


\section{Pilot Method}\label{sec:pilot}

The \emph{pilot method} \cite{Duin1999,Voss2005} or equivalently the \emph{Rollout algorithm} \cite{Bertsekas1997} can enhance any heuristic by a simple look-ahead procedure. The idea is to add one-step look-ahead and hence apply greedy heuristics from different starting points. The procedure is applied repeatedly, effectively building a tree. This procedure is not unlike strategies used in game playing programs, that search a game trees for good moves. In all cases the basic idea is to examine all possible choices with respect to their future advantage. An alternative view is that of a sequential decision problem or dynamic programming problem where a solution is built in stages, whereby the components (in our cases the jobs) are selected one-at-a-time. The first $k$ components form a so called $k$-solution   \cite{Bertsekas1997}. In the same way as a schedule was built in stages in Algorithm~\ref{alg:greedy}, where the $k$-solution is the partial schedule ${\cal S}$. However, for the Pilot method the decisions made at each stage will depend on a look-ahead procedure. The Pilot method is then described in Algorithm \ref{alg:Pilot}. The algorithm may seem a little more complicated than necessary, however, as will be seen in the next section this algorithm is a special case of Monte Carlo tree search. The heuristic rollout is performed $B$ times and each time adding a node to the tree. Clearly if all nodes can be connected to a terminal node, the repetition may be halted before the budget $B$ is reached. This is not shown here for clarity. Furthermore, a new leaf on the tree is chosen such that those closer to the root have priority else branches are chosen arbitrarily with equal probability. In some version of the Pilot method, the tree is not expanded breadth first manner but with some probability allows for depth first search. This would be equivalent to executing line~\ref{line:infty} with some probability. This is also commonly used in MCTS and is called progressive widening.

\begin{algorithm}
\SetAlFnt{\footnotesize}
\SetCommentSty{emph}
\SetKwData{Left}{left}
\SetKwData{This}{this}
\SetKwData{Up}{up}
\SetKwFunction{Rollout}{Rollout}
\SetKwFunction{Simulate}{Simulate}
\SetKwInOut{Input}{input}
\SetKwInOut{Output}{output}
\AlFnt
\Input{Budget $B$, partial sequence ${\cal S}_0$, $\vec{t}=(t_1,\ldots,t_n)$, and heuristic {\bf R}}
\Output{Decision, job to dispatch next $b$}
\BlankLine
$root \leftarrow node$ \tcp*[l]{initialize the root node}
$node.n\leftarrow0, node.\vec{t} \leftarrow \vec{t}, node.child \leftarrow \emptyset$\;
\For{$n\leftarrow 1$ \KwTo $B$}{
${\cal S} \leftarrow {\cal S}_0$ \tcp*{set state to root node state and climb down the tree}
\While{$node.child \ne \emptyset$}{
\For{$j\in J ; node.t_j < m_j$}{
\eIf {$node.n = 0$}{
$Q(j) = \infty$ \nllabel{line:infty}
}{
$Q(j) = U(0,1)$  \tcp*{random value between 0 and 1}\nllabel{line:maxval}
}
}
$j' = \arg\max_{j\in J ; t_j < m_j} Q(j)$ \tcp*{largest $Q$ value, break ties randomly}
${\cal S}\leftarrow {\cal S} \oplus \{(j', \sigma(j',t_{j'}))\}$ \tcp*{dispatch job $j'$}
}
\tcp*[l]{expand node if possible, i.e. ${\cal S}$ is not the complete schedule}
\For{$j\in J ; node.t_j < m_j$}{
$node.child[j].parent \leftarrow node$ \tcp*{keep pointer to parent node}
$node.child[j].child \leftarrow \emptyset$ \tcp*{this node has not been expanded}
$node.child[j].n \leftarrow 0$ \tcp*{and has not been rolled out}
$node.child[j].\vec{t}\leftarrow node.\vec{t}$ \tcp*{copy machine counter from parent node}
$node.child[j].t_j\leftarrow node.t_j+1$ \tcp*{increment machine counter for job}
}
$R=$ \Rollout(${\cal S}, node[{\cal S}].\vec{t}, {\bf R}$) \tcp*{Complete the solution via Pilot heuristic}
\nllabel{line:Rollout}
\Repeat(\it propagate result of rollout up the tree){$node \ne root$}{
$node.n \leftarrow node.n + 1$ \tcp*{number of visits incremented by one}
$node.Q \leftarrow \max(node.Q,R)$ \tcp*{best found solution}
$node \leftarrow node.parent$ \tcp*{climb up the tree to parent node}
}( )
}
$\arg\max_{j\in J ; t_j < m_j}  root.child(j).Q$
\caption{Pilot or rollout algorithm}\label{alg:Pilot}
\end{algorithm}

The greedy algorithm~\ref{alg:greedy} is then used as the Rollout algorithm on line~\ref{line:Rollout}. As will be seen in the following section, the key difference between the MCTS and Pilot method is in the way a node is found to expand in the tree and the manner in which a rollout is performed. Other details of the Algorithm~\ref{alg:Pilot} will also become clearer.

\section{MCTS for Combinatorial Optimization}\label{sec:MCTS}
\subsection{Monte Carlo Tree Search}
Monte-Carlo Tree Search inherits from the so-called Multi-Armed Bandit (MAB) framework \cite{Lai-Robbins}. MAB considers a set of independent $k$ arms, each with a 
different payoff distribution. Here each arm corresponds to selecting a job to be dispatched and the payoff the results returned by a rollout or greedy heuristic. 
Several goals have been considered in the MAB setting; one is to maximize the cumulative payoff gathered
along time ($k$-arm bandit) \cite{Auer02}; another one is to identify the 
arm with maximum payoff (max-$k$ arm) \cite{Cicirello2005,Streeter2006,Rimmel09}. At one extreme is the 
exploitation-only strategy (selecting the arm with best empirical reward); at the
other extreme is the exploration-only strategy (selecting an arm with uniform
probability). 

When it comes to find a sequence of options, the search space is structured as 
a tree\footnote{Actually, the search space may be structured as a graph if  different paths can lead to a same state node. In the context of job-shop scheduling however, 
only a tree-structured search space needs be considered.}. 
In order to find the best sequence, a search tree is iteratively used and extended, 
growing in an asymmetric manner to focus the exploration toward the best regions of the search space.  In each iteration, a tree path a.k.a simulation is constructed through three 
building blocks: the first one is concerned with navigating in the tree; the second one is concerned with extending the tree and assessing the current tree path (reward); the third one updates the tree nodes to account for the reward of the current tree path.


\subsubsection{Descending in the tree}
The search tree is initialized to the root node (current partial schedule). In each given
node until arriving at a leaf, the point is to select among the child nodes
of the current node (Fig. \ref{fig:mcts}, left).  For deterministic optimization problems, the goal is 
to maximize the maximum (rather than the expected) payoff. 
For this aim, a sound strategy has been introduced in \cite{Streeter2006} and used in \cite{Rimmel09}. This approach, referred to as Chernoff rule, estimates the 
upper bound on the maximum payoff of the arm, depending on its number of visits and the maximum value gathered.


However, the main goal of this work is to bridge the gap between the Pilot method and MCTS algorithms. Indeed, the Pilot method, as presented in algorithm~\ref{alg:Pilot}, can be viewed as an MCTS algorithm in which the strategy used to chose next child to explore is to choose the best child after one deterministic rollout using the dispatch rule at hand -- a rather greedy exploitation-oriented strategy. Such strategy is very close to a simple rule to balance exploration and exploitation known in the MCTS world as $\epsilon$-greedy: with probability $1 - \epsilon$, one selects the empirically best child node \footnote{Typically $\varepsilon=0.1$.} (i.e. the one with maximum empirical value); otherwise, another uniformly selected child node is retained. Furthermore, similar to the Pilot method described in the previous section, unexplored nodes (line: \ref{line:infty} in Algorithm\ref{alg:Pilot}) will have priority. However, line: \ref{line:maxval} should be replaced by
$$Q(j) \leftarrow node.child[j].Q$$.

\subsubsection{Extending the tree and evaluating the reward}
\label{sec:MonteCarlo}
\begin{figure}
 \centerline{\includegraphics[width=\textwidth]{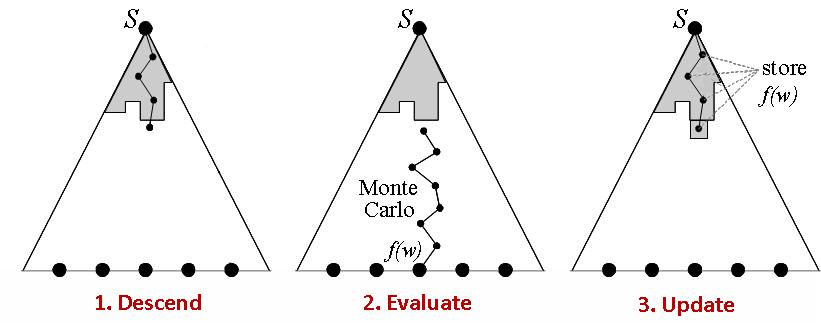}}
\caption{Monte-Carlo Tree Search: the tree is asymmetrically grown toward the most
promising region (in grey, the known part of the tree). Each simulation is made of a MAB phase (1); then the simulation is completed until arriving at a final state (2); finally, the first node of the Monte-Carlo path is added to the known tree, and the reward is computed and back-propagated to all nodes of the path that are in the known tree (3).}
\label{fig:mcts}
\end{figure}

Upon arriving in a leaf, a new option is selected and added as child node
of the current one; the tree is thus augmented of one new node in each simulation
(Fig. \ref{fig:mcts}, right). The simulation is resumed until arriving in a final state (e.g., when all jobs have been processed). As already mentioned, the choices made in the further choice points in the Pilot method rely on the default heuristics (and the rollout is hence deterministic). In the MCTS method however, these choices must rely on randomized heuristics out of consistency with the MAB setting. The question thus becomes which heuristics to use in the so-called random phase (see section \ref{sec:domaink}). Upon arriving in a final state, the reward associated to the simulation is  computed (e.g., the makespan of the current schedule).

\subsubsection{Updating the tree}
The number of visits of the nodes of the current tree that were in the path is incremented by 1; likewise, the 
cumulative reward associated to these nodes is incremented by the reward 
associated to the current path. Note that other statistics can be maintained in each node, such as the RAVE values \cite{Gelly07}, and must be updated there too.

\subsection{Monte-Carlo Tree Schedule (\MCS)}\label{sec:domaink}

As already pointed out, solving a combinatorial problem can be viewed as a sequential decision making process, incrementally building a solution by choosing an element of a partial solution at a time (e.g., next town for TSP problems, of next machine to schedule for the job shop scheduling problem). In order to 
solve this sequential decision problem through a MCTS algorithm, several specific issues must be considered.

A first issue concerning the reward design has a significant impact on 
the exploration {\em versus} exploitation dilemma; it might require some instance-dependent parameter to reach a proper balance (see e.g., \cite{banditLION09}). Indeed, in the case of job-shop scheduling for instance, different instances will have very different makespans. 

A second issue concerns the heuristics to be used in the random phase (section 
\ref{sec:MonteCarlo}). The original MCTS method \cite{Kocsis06} advocates pure random choices. Domain knowledge could however be used to propose a smarter
procedure, e.g. using random selected dispatching rules for job-shop scheduling. 
Still, a lesson consistently learned from MCTS applications \cite{Gelly07,Rimmel09} is that doing many simulations completed with a brute random phase is more effective than doing less simulations completed with a smart final phase. For instance in the domain of computer-Go, the overall results were degraded by using 
gnuGo in the random phase, compared to a uniform move selection.
Likewise, the use of the three dispatching rules (described in Section \ref{sec:scheduling}) in the random phase was outperformed by a pure random strategy, uniformly selecting the next job to be considered.
Here we have chosen to follow this path and our rollout phase consists of the purely random dispatching of jobs. Replacing this with line \ref{line:Rollout} in Algorithm 2, along with the $\varepsilon$--greedy policy, and the Pilot method is transformed into MCTS. 

Another important detail must be taken into account: When performing random rollouts, we might forget the best choices found previously during the building of a schedule. For this reason, a global best found sequence is kept throughout the scheduling procedure. If a suboptimal choice is found in a later partial schedule ($k$-solution), the globally better choice previously found during a random rollout will be forced. In some sense the idea here is similar to that of the fortified rollouts used by the Pilot method \cite{Voss2005}.

Finally, in MCTS, the stopping criterion is defined by the total number of simulations (i.e., rollouts here), and one single decision is taken after a complete tree exploration, chosen as the child of the root node with the maximum expected reward \cite{Gelly07}. The situation is rather different here, and, in this first approach to MCTS for combinatorial optimization, similarly to the Pilot method, a single tree exploration is done, with a limited budget in terms of number of rollouts. A complete schedule is then built by descending the tree and always choose the child with maximum expected reward (makespan). 

\section{Experimental Results}
\label{sec:experiments}
This section reports on the experimental validation of the proposed
approaches.  After detailing the experimental setting, the results of
the \MCS\ method is reported and compared to the baseline methods, the greedy application of the three
dispatching rules MWKR, SPT and LOPN (section \ref{sec:scheduling}),
and the pilot methods built on these three dispatching rules (section \ref{sec:pilot}).

\subsection{Experimental Settings}
\label{sec:conditions}

All experiments have been conducted using the set of test instances proposed in \cite{Taillard1993}.
The machine orders for the jobs are randomly generated and the processing times are discrete values uniformly distributed between 1 and 200. Three different $n\times m$ problem sizes were generated using this setup, $6\times 6$, $10\times 10$ and $14\times 14$. The optimal makespans for one hundred instances generated of each size was then found using Brucker's branch and bound algorithm \cite{Brucker2007}. A further four instance of size $20 \times 20$ are also tested \cite{Yamada1992} and compared with their best known solution \cite{Banharnsakun2011}.

For all methods except the greedy ones, the time budget is varied to assess the 
convergence behavior of the pilot and \MCS\ optimization methods, considering a total of 100, 1,000 and 5,000 rollouts a.k.a. simulation. Each rollout corresponds
to designing and evaluating a complete solution (computing its total makespan).
It is worth noting that not all rollouts are equally expensive; the rollout based on a dispatching rule (as used in the pilot methods) is more computationally demanding than the random rollout used in \MCS, all the more so as the size of the problem instance increases. Nevertheless, the fixed rollout budget is meant to 
allow CPU-independent comparisons and assess the empirical behavior of the methods
under restricted computational resources (e.g. in real-world situations). 


For each method, each problem size and each time budget, 
the result is given as the average
over 100 problem instances of the normalized makespan (1. being the optimal value), 
together with the minimum, maximum and median values, and the standard deviation;
the number of times where the optimal value was found is additionally reported.

While the greedy and pilot algorithms actually are deterministic\footnote{Up to 
ties between jobs.}, \MCS\ is not. The usual way to measure the performance of a stochastic algorithm on a given problem domain is through averaging the result out of a few dozen or hundred independent runs. For the sake of computational convenience however, \MCS\ was run only once
on each problem instance and the reported result is the average over the 100 
independent instances.

\subsection{The Greedy Algorithms}
\begin{table}[t!]
\label{resultsDispatchRules}
\begin{center}
\begin{tabular}{l|l|ccccc|c}
Instance & Heuristic & min & mean & median & stdev & max & \#opt\\\hline
$ 6\times  6$ & MWKR & 1.000 & 1.155 & 1.151 & 0.084 & 1.384 & 2 \\
& SPT  & 1.137 & 1.399 & 1.390 & 0.150 & 1.816 & 0 \\
& LOPN & 1.017 & 1.176 & 1.178 & 0.083 & 1.369 & 0 \\\hline\hline
$10\times 10$ & MWKR & 1.096 & 1.228 & 1.222 & 0.069 & 1.430 & 0 \\
& SPT  & 1.303 & 1.654 & 1.644 & 0.166 & 2.161 & 0 \\
& LOPN & 1.103 & 1.216 & 1.208 & 0.061 & 1.369 & 0 \\\hline\hline
$14\times 14$  & MWKR & 1.159 & 1.264 & 1.261 & 0.052 & 1.399 & 0 \\
& SPT  & 1.584 & 2.012 & 2.015 & 0.244 & 2.721 & 0 \\
& LOPN & 1.150 & 1.253 & 1.250 & 0.048 & 1.376 & 0 \\
\end{tabular} 
\end{center}
\caption{Greedy algorithms: Performance statistics of the MWKR, SPT and LOPN rules on three problem sizes. }\label{tbl:pureheuristic}
\end{table}

The results of the greedy algorithms are depicted in Table \ref{tbl:pureheuristic} for the three dispatching rules MWKR, SPT and LOPN (section \ref{sec:scheduling}), showing that MWKR and LOPN behave similarly and
significantly outperform SPT. Further, the performances of SPT significantly decrease with the problem size, whereas MWKR and LOPN demonstrate an excellent
scalability in the considered size range.

\begin{table}[tb!]
 {\footnotesize
 \begin{center}
\begin{tabular}{l|l|ccccc|c}
Instance & Heuristic & min & mean & median & stdev & max & \#opt\\\hline
$ 6\times  6$ & Pilot(MWKR,100) & 1.000 & 1.049 & 1.050 & 0.038 & 1.167 & 14 \\
& Pilot(MWKR,1000) & 1.000 & 1.035 & 1.030 & 0.034 & 1.180 & 23 \\
& Pilot(MWKR,5000) & 1.000 & 1.025 & 1.014 & 0.029 & 1.104 & 33 \\\hline
& Pilot(SPT,100) & 1.000 & 1.100 & 1.093 & 0.066 & 1.293 & 3\\
& Pilot(SPT,1000) & 1.000 & 1.065 & 1.060 & 0.050 & 1.287 & 7\\
& Pilot(SPT,5000) & 1.000 & 1.052 & 1.049 & 0.045 & 1.265 & 14\\\hline
& Pilot(LOPN,100) & 1.000 & 1.058 & 1.057 & 0.044 & 1.172 & 12\\
& Pilot(LOPN,1000) & 1.000 & 1.046 & 1.036 & 0.040 & 1.134 & 17\\
& Pilot(LOPN,5000) & 1.000 & 1.034 & 1.024 & 0.032 & 1.127 & 22\\\hline\hline
$10\times 10$ & Pilot(MWKR,100) & 1.032 & 1.109 & 1.109 & 0.039 & 1.217 & 0\\
& Pilot(MWKR,1000) & 1.006 & 1.097 & 1.096 & 0.039 & 1.215 & 0\\
& Pilot(MWKR,5000) & 1.004 & 1.082 & 1.083 & 0.035 & 1.158 & 0\\\hline
& Pilot(SPT,100) & 1.102 & 1.222 & 1.221 & 0.063 & 1.427 & 0\\
& Pilot(SPT,1000) & 1.066 & 1.188 & 1.188 & 0.055 & 1.332 & 0\\
& Pilot(SPT,5000) & 1.063 & 1.172 & 1.168 & 0.048 & 1.296 & 0\\\hline
& Pilot(LOPN,100) & 1.044 & 1.117 & 1.114 & 0.041 & 1.219 & 0\\
& Pilot(LOPN,1000) & 1.028 & 1.106 & 1.105 & 0.042 & 1.212 & 0\\
& Pilot(LOPN,5000) & 1.022 & 1.096 & 1.092 & 0.032 & 1.171 & 0\\\hline\hline
$14\times 14$ & Pilot(MWKR,100) & 1.081 & 1.156 & 1.155 & 0.036 & 1.256 & 0\\
& Pilot(MWKR,1000) & 1.046 & 1.142 & 1.138 & 0.036 & 1.247 & 0\\
& Pilot(MWKR,5000) & 1.046 & 1.129 & 1.129 & 0.034 & 1.230 & 0\\\hline
& Pilot(SPT,100) & 1.239 & 1.389 & 1.380 & 0.080 & 1.595 & 0\\
& Pilot(SPT,1000) & 1.136 & 1.316 & 1.319 & 0.065 & 1.508 & 0\\
& Pilot(SPT,5000) & 1.153 & 1.286 & 1.283 & 0.060 & 1.517 & 0\\\hline
& Pilot(LOPN,100) & 1.076 & 1.160 & 1.161 & 0.036 & 1.285 & 0 \\
& Pilot(LOPN,1000) & 1.080 & 1.149 & 1.152 & 0.037 & 1.264 & 0 \\
& Pilot(LOPN,5000) & 1.078 & 1.145 & 1.142 & 0.033 & 1.248 & 0 \\

\end{tabular}
 \end{center}}
 \caption{Pilot algorithms. Performance statistics for Pilot algorithm using 3 different Pilot heuristics: MWKR, SPR, and LOPN, on the three different problem sizes.}
 \label{tbl:stats_pilot}
\end{table}

\subsection{The Pilot Method}
The results of the pilot method (section \ref{sec:pilot}) related to the three above dispatching rules and three time budgets (100, 1000 and 5000) are displayed in table \ref{tbl:stats_pilot} for all three problem sizes $6 \times 6$, $10 \times 10$, and $14 \times 14$. For the sake of easy comparison with the greedy algorithm, the median performances respectively obtained on the same problem sizes displayed on Table \ref{tbl:pilotgreedy}.


As was expected, and demonstrated on some TSP instances by \cite{Duin1999}, 
the pilot method does improve on the greedy algorithm.
With a time budget 100, a significant improvement is
observed for all three methods, and confirmed by the number of times the 
optimal solution is discovered on problem size 6 x 6. 

It is worth noting that the performance only very slightly improves when the time budget increases from 100 to 1000, and from 1000 to 5000, despite the significant increase in the computational effort. In particular, the optimal solutions are never discovered for higher problem sizes. 

Lastly, the performance order of the 
three rules is not modified when using the pilot method: MWKR and LOPN significantly outperform SPT. As a consequence the Pilot method can be quite sensitive to the Pilot heuristic chosen.

\begin{table}[htbp]
\begin{center}
\begin{tabular}{l|l|cccc}
Instance & Heuristic & \multicolumn{4}{c}{Algorithm (budget)} \\\hline
~ & ~ & ~ & Pilot (100) & Pilot (1000) & Pilot (5000)  \\\hline
$ 6\times  6$ & MWKR & 1.151 & 1.050 & 1.030 & 1.014\\
& SPT  & 1.390 & 1.093 & 1.060 & 1.049 \\
& LOPN & 1.178 & 1.057 & 1.036 & 1.024\\\hline\hline
$10\times 10$ & MWKR & 1.222 & 1.109 & 1.096 & 1.083\\
& SPT  & 1.644 & 1.221 & 1.188 & 1.168  \\
& LOPN & 1.208 & 1.114 & 1.105 & 1.092 \\\hline\hline
$14\times 14$  & MWKR & 1.261 & 1.155 & 1.138 & 1.129\\
& SPT  & 2.015 & 1.380 & 1.319 & 1.283\\
& LOPN & 1.250 & 1.161 & 1.152 & 1.142\\
\end{tabular} 
\end{center}
\caption{Comparison of median performances of Greedy and Pilot with different budgets, for the 3 heuristics (from Tables \ref{tbl:pureheuristic} and \ref{tbl:stats_pilot}).}\label{tbl:pilotgreedy}
\end{table}

\subsection{Monte-Carlo Tree Schedule}
Table \ref{tbl:stats_bandit} reports on the results of the \MCS\ approach described in (section \ref{sec:MCTS}). 
On problem size 6x6, \MCS\ significantly improves on the best Pilot
method (MKWR with 5000 rollout budget), as also witnessed by the
number of times the optimal solution is found. On problem size 10x10,
the average and median performances are comparable; still, the
optimal solution is found twice by \MCS\ with a 5000 rollout budget,
whereas it is never found by
the Pilot method. On problem size 14x14, Pilot (MWKR,5000) is
significantly better than \MCS. A first explanation for this fact
relies on the computational effort: As already mentioned, the
computational time required for a 5,000 rollout Pilot is circa 4 times
higher than for a 5,000 rollout \MCS.  A second explanation is 
the fact that, as the tree depth increases with the problem size, it
becomes necessary to adjust the parameters controlling the branching
factor of the \MCS\ tree. This can be achieved by introducing progressive widening (on-going work).

\begin{table}[t!]
 {
 \begin{center}
\begin{tabular}{l|l|ccccc|c}
Instance & Heuristic & min & mean & median & stdev & max & \#opt\\\hline
$ 6\times  6$ & & &  & & & &\\
& \MCS($\varepsilon$--greedy,100) & 1.000 & 1.026 & 1.020 & 0.027 & 1.097 & 28\\
& \MCS($\varepsilon$--greedy,1000) & 1.000 & 1.014 & 1.001 & 0.025 & 1.150 & 50\\
& \MCS($\varepsilon$--greedy,5000) & 1.000 & 1.007 & 1.000 & 0.017 & 1.082 & 72\\\hline\hline
$10\times 10$ & & &  & & & &\\
& \MCS($\varepsilon$--greedy,100) & 1.029 & 1.141 & 1.140 & 0.057 & 1.378 & 0\\
& \MCS($\varepsilon$--greedy,1000) & 1.014 & 1.095 & 1.098 & 0.036 & 1.199 & 0\\
& \MCS($\varepsilon$--greedy,5000) & 1.000 & 1.070 & 1.070 & 0.032 & 1.135 & 2\\\hline\hline
$14\times 14$ & & &  & & & & \\
& \MCS($\varepsilon$--greedy,100) & 1.173 & 1.346 & 1.331 & 0.083 & 1.634 & 0 \\
& \MCS($\varepsilon$--greedy,1000) & 1.127 & 1.284 & 1.280 & 0.074 & 1.552 & 0 \\
& \MCS($\varepsilon$--greedy,5000) & 1.064 & 1.232 & 1.223 & 0.065 & 1.471 & 0 \\
\end{tabular}
 \end{center}}
\caption{Monte-Carlo Tree Schedule. Performance statistics using $\varepsilon$--greedy and random scheduling.}
 \label{tbl:stats_bandit}
\end{table}

\subsection{Larger Instances}

The best known solutions (BKS) for the $20\times 20$ benchmark problem \cite{Yamada1992} are taken from \cite{Banharnsakun2011}. Here we demonstrate the performance of the \MCS\ on problems that cannot be solved using exact methods. Only four instances are considered here, and the \MCS\ is run 30 times on each instance. The method is unable to find any best known solution. Nevertheless, the performance does not degrade significantly when compared to the results obtained on the $10 \times 10$ problems.

\begin{table}[t!]
 {\footnotesize
 \begin{center}
\begin{tabular}{l|l|ccccc|c}
Instance & Heuristic & min & mean & median & stdev & max & \#BKS\\\hline
 yn01 & \MCS($\varepsilon$--greedy,100) & 1.166 & 1.211 & 1.210 & 0.020 & 1.245 & 0 \\
  & \MCS($\varepsilon$--greedy,1000) & 1.144 & 1.186 & 1.183 & 0.021 & 1.235 & 0 \\
  & \MCS($\varepsilon$--greedy,5000) & 1.128 & 1.167 & 1.166 & 0.023 & 1.209 & 0 \\\hline
 yn02 & \MCS($\varepsilon$--greedy,100) & 1.135 & 1.185 & 1.181 & 0.024 & 1.228 & 0 \\
  & \MCS($\varepsilon$--greedy,1000) & 1.123 & 1.156 & 1.153 & 0.023 & 1.218 & 0 \\
  & \MCS($\varepsilon$--greedy,5000) & 1.090 & 1.130 & 1.130 & 0.023 & 1.174 & 0 \\\hline
 yn03 & \MCS($\varepsilon$--greedy,100) & 1.156 & 1.205 & 1.204 & 0.022 & 1.267 & 0 \\
  & \MCS($\varepsilon$--greedy,1000) & 1.131 & 1.178 & 1.180 & 0.020 & 1.222 & 0 \\
  & \MCS($\varepsilon$--greedy,5000) & 1.110 & 1.151 & 1.148 & 0.023 & 1.199 & 0 \\\hline
 yn04 & \MCS($\varepsilon$--greedy,100) & 1.063 & 1.108 & 1.107 & 0.025 & 1.160 & 0 \\\
  & \MCS($\varepsilon$--greedy,1000) & 1.039 & 1.084 & 1.090 & 0.025 & 1.133 & 0 \\\
  & \MCS($\varepsilon$--greedy,5000) & 1.023 & 1.067 & 1.064 & 0.019 & 1.102 & 0 \\

\end{tabular}
 \end{center}}
 \caption{Results for the \MCS\ on four $20 \times 20$ instances. Performance statistics is given for 30 independent runs on each instance, yn01,$\ldots,$ yn04.}
 \label{tbl:stats_yamada}
\end{table}

\section{Discussion and Perspectives}

The main contribution of this paper is to demonstrate the feasibility
of using MCTS to address job-shop scheduling problems. This result
has been obtained by using the simplest exploration/exploitation strategy 
in MCTS, the $\varepsilon$-greedy strategy, defining the {\em Monte-Carlo
   Tree Scheduling} approach (\MCS). The empirical evidence gathered
 from the preliminary experiments presented here shows that \MCS\ significantly
 outperforms its competitors on small and medium size problems. For
 larger problem sizes however, the Pilot method with the best
 dispatching rule outperforms this first verions of \MCS. This fact is blamed on our
 adversary experimental setting, as we compared methods based on the 
 number of rollouts, whereas the computational cost of a rollout
 is larger by almost an order of magnitude in the Pilot framework, as
 compared to that of the \MCS.
 
The \MCS\ scalability can also be improved through reconsidering the exploration vs 
exploitation trade-off, ever more critical in larger-sized problem instances.
First of all, the MAX-k-arm strategy should be tried in lieu of the simple $\varepsilon$-greedy rule.
Furthermore, this tradeoff can be also adjusted by avoiding the systematic first trial of all possible children, as this becomes harmful for large number or arms (jobs here).
It is possible to control when a new child node 
 should be added in the tree, and which one. Regarding the former aspect, 
 a heuristics referred
 to as {\em Progressive Widening} has been designed to limit the branching factor
 of the tree, e.g. \cite{Rolet09} . Regarding the second aspect, 
 the use of a Rapid Action Value Estimate (RAVE), 
 first developed in the computer-Go context \cite{Gelly07} can be very efficient
 to aggregate the various rewards computed for the same option ({\em Queen 
 Elisabeth}), and guide the introduction of the most efficient rules/jobs 
 in average.

\section{Conclusion and outlook}

This work has shown how the Pilot method may be considered a special case of MCTS, with an exploratory-only strategy to traversing the tree and using a deterministic rollout driven by the Pilot heuristic. It has demonstrated that
the Pilot method can be sensitive to the chosen Pilot heuristic. As the chosen Pilot heuristic becomes more effective, so too may its computational costs. An extension of the Pilot method in the realm of MCTS algorithms, the \MCS, has been proposed, using a simple $\epsilon$--greedy strategy to traverse down the tree. However, more sophisticated strategies, such as the one based on the max-$k$ bandit problem \cite{Streeter2006,Cicirello2005}, need now be investigated. For larger problems, progressive widening should be an avenue for further research, as similar strategies have already been investigated in the Pilot framework. Finally, Rapid Action Value Estimates may not only be used to bias how the tree is traversed, possibly replacing the exploration term in the bandit formulas, but can also help to improve over the random rollouts.


\end{document}